# Synchronous Counter Design Using Novel Level Sensitive T-FF in QCA Technology

**Ali H. Majeed** [1,2], **Esam Alkaldy** [2,*], **Mohd Shamian bin Zainal** [1] **and Danial Bin MD Nor** [1]

[1] Faculty of Electrical and Electronic Engineering, UTHM, Johor 86400, Malaysia
[2] Electrical Department, College of Engineering, University of Kufa, Kufa 54003, Iraq
* Correspondence: esam.alkaldy@uokufa.edu.iq

**Abstract:** The quantum-dot cellular automata (QCA) nano-technique has attracted computer scientists due to its noticeable features such as low power consumption and small size. Many papers have been published in the literature about the utilization of this technology for de-signing many QCA circuits and for presenting logic gates in an optimal structure. The T flip-flop, which is an essential part of digital designs, can be used to design synchronous and asynchronous counters. This paper presents a novel T flip-flop structure in an optimal form. The presented novel gate was used to design an N-bit binary synchronous counter. The QCADesigner software was used to verify the designed circuits and to present the simulation results, while the QCAPro tool was used for the power analysis. The proposed design required minimal power and showed good improvements over previous designs.

## 1. Introduction

Quantum-dot cellular automata (QCA) is one of the new nanoelectronics that has emerged in the last decade. QCA technology was first introduced by Lent et al. in 1993 [1]. QCA is being used as a new technique for computation. QCA is a good choice as a replacement for CMOS technology due to many aspects such as its ultra-high speed, small size, and low power consumption. It depends on electron configurations instead of voltage levels as in CMOS. The key issue in QCA is complexity reduction. Many papers have been presented on designing important digital circuits using QCA technology. Most of these papers were searching for an optimal form. Researchers have been paying attention to the design of the memory cell as one of the important circuits in QCA technology [2–7], and a well-optimized flip-flop structure [8–10], in addition to counter circuits [9–11]. This paper introduced an optimal T flip-flop structure with a super degradation in cell counts and used it to design highly efficient N-bit counter circuits. The presented flip-flop was used to carry out many counter circuits.

In QCA technology, researchers look for optimization in terms of area, delay, and cells required. As such, optimization of the brick units is a very essential issue. In this paper, T flip-flop represents the selected domain while the counter circuit represents the case study to show the power of the proposed gate.

The rest of this work is as follows: Section 2 focuses on the QCA fundamentals, Section 3 on previous works, Section 4 on the proposed design, Section 5 provides the simulation results with comparison tables, Section 6 gives an analysis of the dissipated power for the proposed flip-flop, and finally, the conclusion of this paper is presented in Section 7.

## 2. QCA Fundamentals

A quantum cell is a brick unit in QCA. Each square-shaped cell consists of four dots distributed regularly inside the cell. Each cell has two electrons that can tunnel between the dots but cannot escape out of the cell. The electron repulsion forces the electrons to occupy the dots diagonally. The construction of the electrons inside the cell can represent the digital binary numbers; where the cell polarization (P) equals +1, it is represented as binary 1; while when P equals −1, it is represented as binary 0, as in Equation 1. Figure 1 shows the state of the cell at different polarizations.

$$P = \frac{(p1+p3)-(p2+p4)}{\sum_{i=1}^{4} pi} \quad (1)$$

where if the dot has an electron p = 1, elsewhere p = 0.

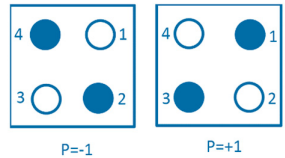

**Figure 1.** Cell polarizations.

The main building blocks in QCA are the 3-input majority gate (Maj-3), as illustrated in Figure 2, and the inverter shown in Figure 3. Any Boolean function can be represented by using these gates. The AND gate can be formed using the majority gate by connecting one of its inputs to logic 0. If any inputs of the majority gate connect to logic 1, the OR gate will be obtained. The logical function of the majority gate is illustrated in Equation 2.

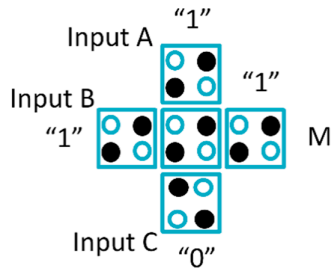

**Figure 2.** Majority gate.

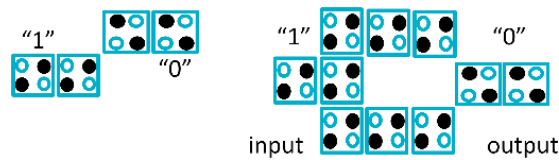

**Figure 3.** Quantum-dot cellular automata (QCA) inverter forms.

$$\text{Maj}(X, Y, Z) = XY + XZ + YZ \quad \ldots \quad (2)$$

The multi-input majority was considered by researchers [12–16], and its reliability was studied in [17].

The clock is a vital issue in QCA for the following reasons: The clock signal gives the circuit the ability to make the synchronization, and it controls the direction of the data flow. Clocking can be considered as the major source of power to stimulate the QCA circuit. QCA circuits use four clocking signals to achieve the inherent pipelining, with each signal having four phases, as illustrated in Figure 4. The clock phases provide adiabatic switching instead of abrupt switching for the cells, thereby making the circuit closer to the ground state [18,19].

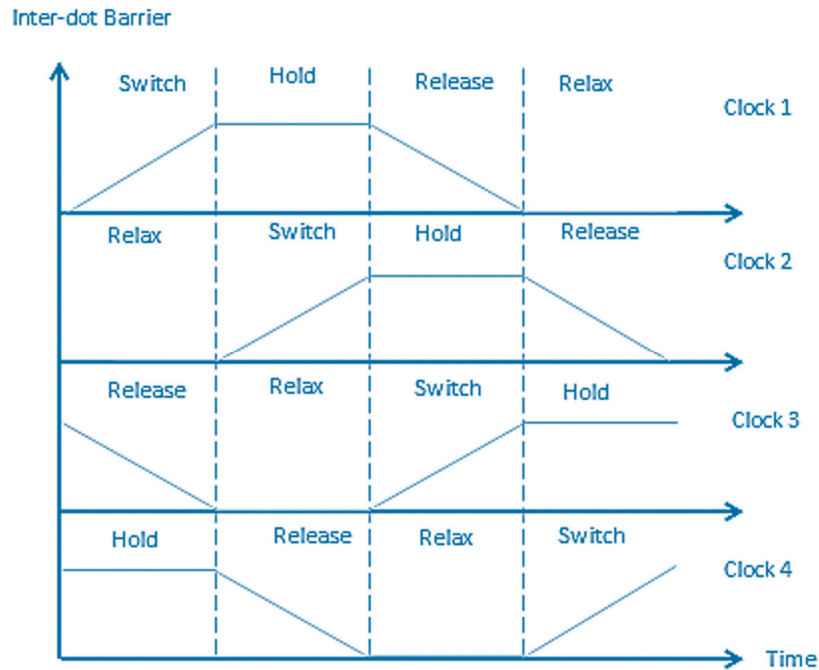

**Figure 4.** QCA clock signals.

## 3. Previous Works

A counter is a type of sequential circuit constructed with a set of flip-flops connected in a suitable manner to be able to count sequence of inputted pulses. The counters are categorized into two types depending on the topology that is connected; asynchronous and synchronous. In the asynchronous counter, the output signal of one flip-flop represents a clock to the next one. As such, the time delay in this type of counter is the sum of all flip-flop's propagation delay. This drawback is overcome by the synchronous counter. In the synchronous counter, all flip-flops are connected to the same clock signal. As such, the time delay in this type of counter is the same propagation delay as a single flip-flop. The synchronous counter is widely used in digital systems. It can count from 0–$2N–1$, where N is the number of the flip-flops used, or counter size. This paper introduces a novel structure of the T flip-flop because the T flip-flop is widely used in digital counters, binary addition, and frequency divider circuits. As such, the designers looking for the optimal structure of the QCA flip-flop to obtain an optimal N-bit counter circuit [9–11,20]. Figure 5 illustrates some of the T flip-flops presented previously.

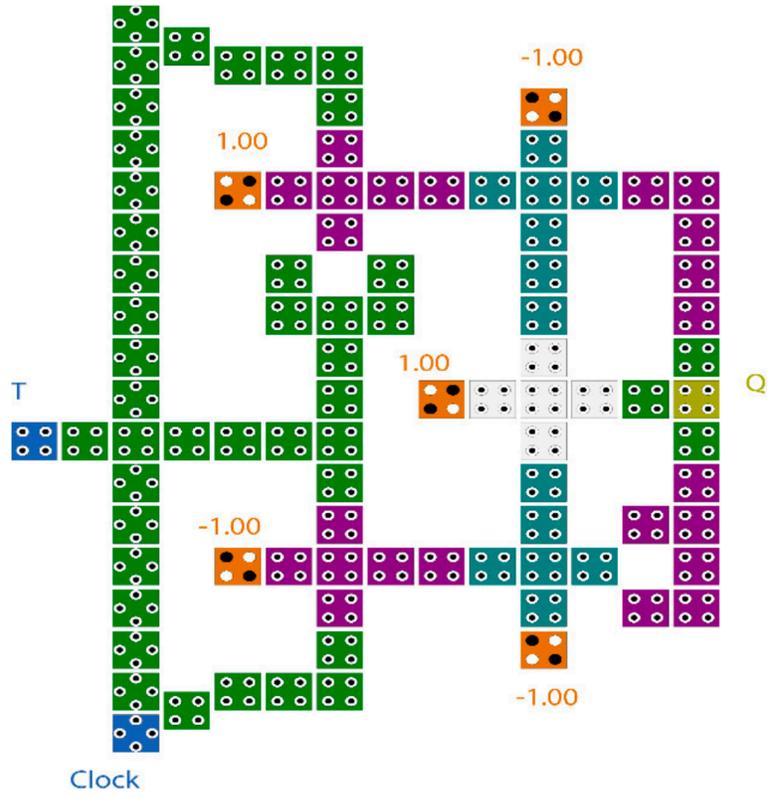

(**a**)

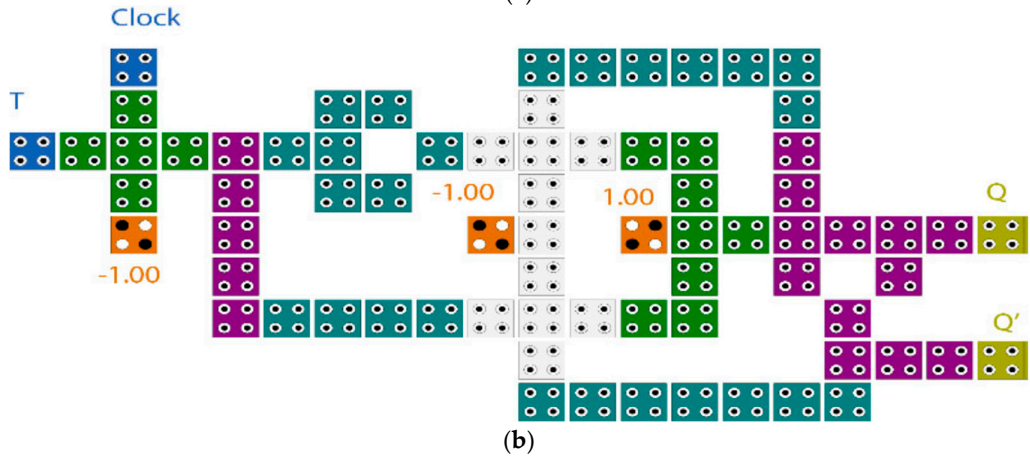

(**b**)

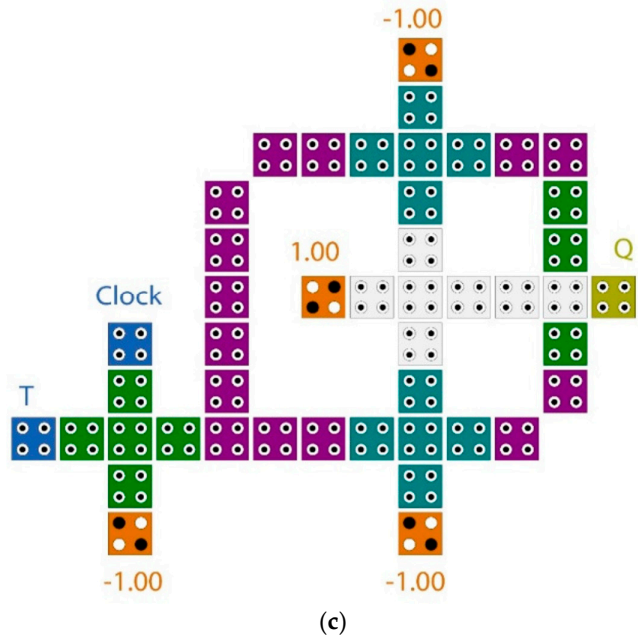

(**c**)

**Figure 5.** Previous designs of QCA T flip-flops (T-FF) presented in (**a**) [21] (**b**) [22] (**c**) [10].

This paper presented a new and efficient T flip-flop (T-FF) using a block diagram designed to optimize the synchronous counter in terms of area and complexity. In addition, a new falling edge converter was used for the presented circuit.

## 4. Proposed Design

In this section, a new T-FF structure will be introduced, and then, by using the appropriate converters, the basic elements will be created to carry out a QCA N-bit synchronous counter. The presented flip-flop was a level-sensitive design.

*4.1. T Flip-Flop*

The schematic diagram of the proposed T flip-flop is illustrated in Figure 6a, while the QCA layout is shown in Figure 6b. It is clear from the presented diagram that the T flip-flop was accomplished by using the XOR gate with the AND gate. The structure of the XOR gate used in this paper was introduced by [23]. The data storage was accomplished using a loop-based mechanism. Figure 7 illustrates the electron configurations inside cells for many cases of input and previous output. The proposed T flip-flop was level-sensitive and was constructed with only 21 cells with an area of 0.018 $\mu m^2$. It was evident that the first significant waveform was obtained at the output after one clock cycle at clock phase 2, which was one clock phase faster than the best previously introduced counterparts. If the input signal (T) is 1 when the clock is available at a high level, the output will flip to the inverse state. Otherwise, the output will remain unchanged. The level-sensitive T flip-flop (LST-FF) presented in this paper had an improvement of up to 54% over the best designs available. The truth table of the proposed flip-flop is detailed in Table 1.

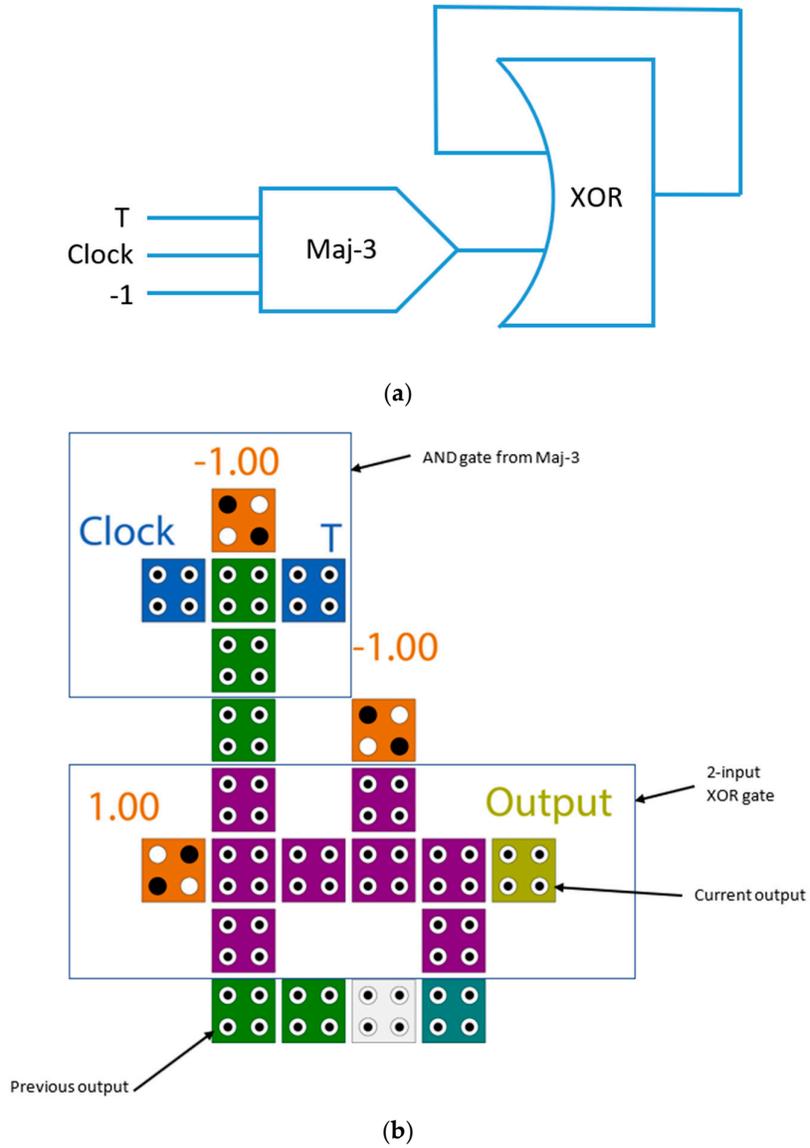

**Figure 6.** The proposed level-sensitive T flip-flop (LST-FF) (**a**) schematic diagram (**b**) QCA layout.

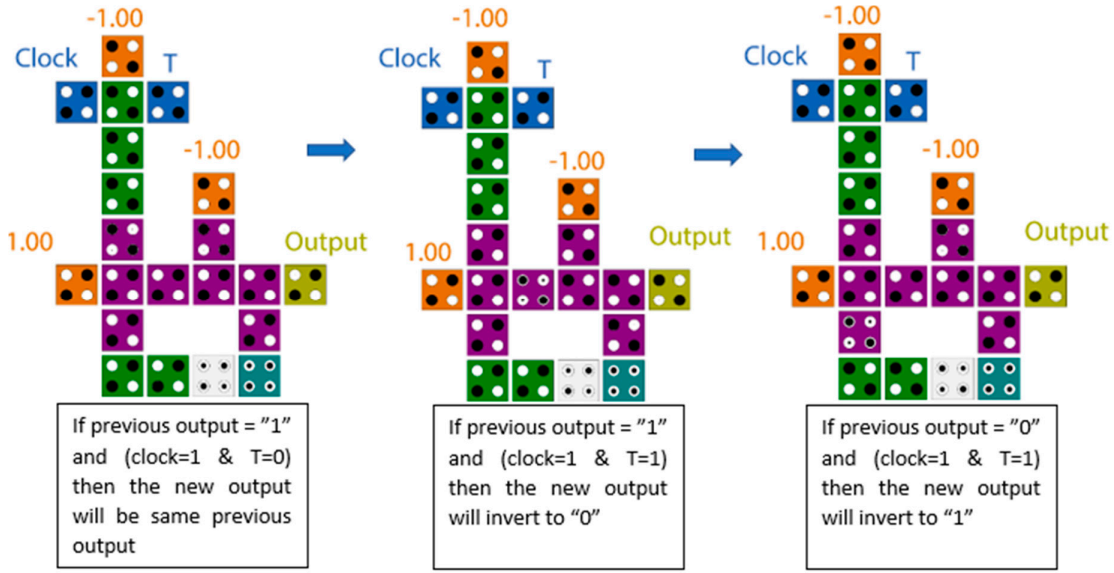

**Figure 7.** Electrons configurations for many cases of input and previous output.

**Table 1.** Proposed LST-FF truth table.

| T | Clock | Current Output ($Q_t$) |
|---|---|---|
| 0 | 0 | $Q_{t-1}$ |
| 0 | 1 | $Q_{t-1}$ |
| 1 | 0 | $Q_{t-1}$ |
| 1 | 1 | $Q_{t-1}$ |

*4.2. Synchronous Counter*

The novel LST-FF that was introduced in this paper was utilized to design a new layout synchronous counter. In this work, the 3-bit counter circuit illustrated in Figure 8 was introduced as an example of an N-bit counter. The proposed counter used an optimal QCA circuit to accomplish a negative edge trigger. This edge detection circuit, as illustrated in Figure 7, was constructed with an AND gate and inverter, and gave a high output only by the transition of 1–0, as demonstrated in Table 2.

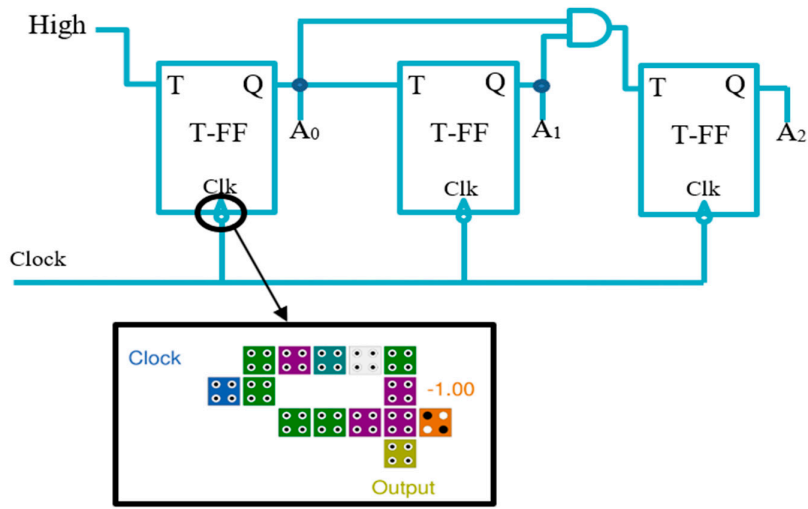

**Figure 8.** Synchronous 3-bit counter with negative edge-triggered QCA circuit.

**Table 2.** Negative edge trigger functionality table.

| Previous Clock | Current Clock | Output |
|---|---|---|
| 0 | 0 | 0 |
| 1 | 0 | 1 |
| 0 | 1 | 0 |
| 1 | 1 | 0 |

This paper implemented a 3-bit QCA synchronous counter circuit, as illustrated in Figure 9. This mod 8 counter was able to count sequentially from 0 to 7 in decimal mode. The three output waveforms (A2, A1, and A0) were combined to trace the order from 000 to 111, consecutively.

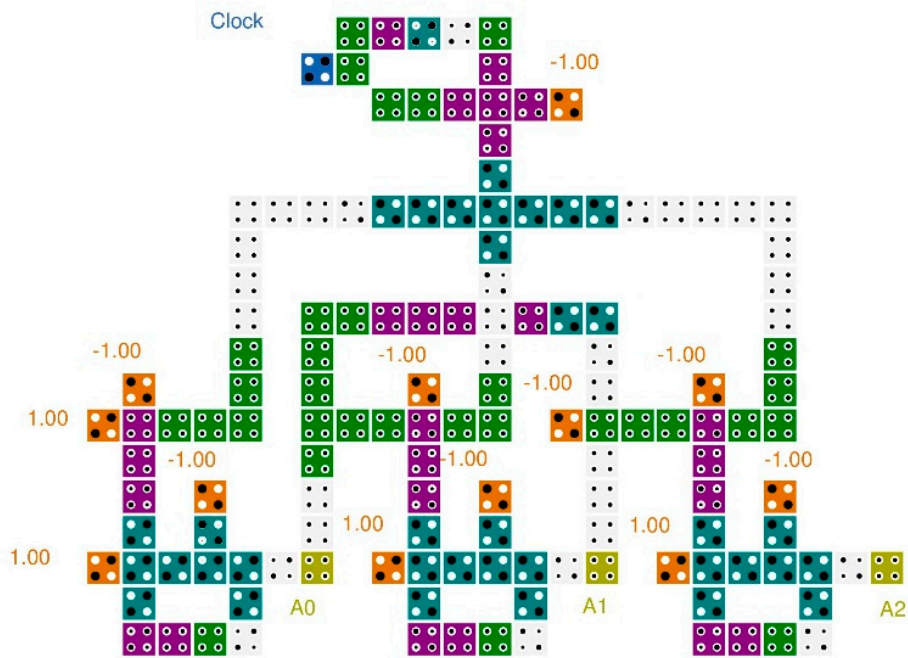

**Figure 9.** The proposed QCA-circuit of the 3-bit synchronous counter.

## 5. Simulation Results and Comparison

This section explains the output waveforms of the proposed designs of the LST-FF and 3-bit synchronous counter as well as the tables of comparisons with previous counterparts. This paper evaluated N-bit counters (up to 4). The output waveforms of the 3-bit counter are given as examples, and the details of the other counters are listed in the comparison results shown in Table 4. Figure 10 illustrates the simulation output of the proposed LST-FF, while the output waveforms of the presented 3-bit counter are illustrated in Figure 11.

**Figure 10.** Output waveforms of the proposed LST-FF.

**Figure 11.** Output waveforms of the presented 3-bit synchronous counter.

The novel LST-FF consisted of only 21 cells and had a noticeably small area of 0.0186 μm². The reduced complexity of this flip-flop with the best reported design was 69% and 54% in terms of the area and number of cells required, respectively, as illustrated in Table 3. The 3-bit single-layer synchronous counter design used in this work had a complexity reduction of 20% and 19.5% when compared with the best previous design in terms of area and cell counts, respectively, as detailed in Table 4.

Table 3. Comparison of results for the proposed LST-FF.

| Design | Cell counts | Area (μm²) | Latency |
|---|---|---|---|
| [24] | 184 | 0.32 | 3 |
| [25] | 108 | 0.20 | 1.5 |
| [21] | 92 | 0.10 | 1.25 |
| [26] | 81 | 0.07 | 1.5 |
| [22] | 66 | 0.06 | 1.25 |
| [27] | 55 | 0.06 | 1.5 |
| [10] | 46 | 0.06 | 1 |
| Proposed | 21 | 0.0186 | 0.5 |

Table 4. Comparison of results for the presented counters.

| Design | No. of Bit | Cell Counts | Area (μm²) | latency | Layer Required |
|---|---|---|---|---|---|
| [9] | 2 | 328 | 0.62 | 3 | Single |
| | 3 | 616 | 1.2 | 5 | Single |
| | 4 | 1130 | 2.2 | 7 | Single |
| [11] | 2 | 240 | 0.26 | 2 | Multi |
| | 3 | 428 | 0.48 | 2 | Multi |
| | 4 | 652 | 0.74 | 2 | Multi |
| [10] | 2 | 141 | 0.22 | 2.25 | Single |
| | 3 | 238 | 0.36 | 2.25 | Single |
| | 4 | 354 | 0.49 | 2.25 | Single |
| [28] | 2 | - | - | - | - |
| | 3 | 196 | 0.22 | 4 | Single |
| | 4 | - | - | - | - |
| [29] | 2 | - | - | - | - |
| | 3 | 174 | 0.20 | 3 | single |
| | 4 | 258 | 0.25 | 4 | single |
| Proposed | 2 | 80 | 0.09 | 2 | Single |
| | 3 | 140 | 0.16 | 2 | Single |
| | 4 | 196 | 0.24 | 2 | Single |

## 6. Power Analysis

The power dissipation of the proposed LST-FF was estimated using the QCAPro tool. This tool is capable of dealing with a large number of cells because it utilizes a fast approximation-based technique, and non-adiabatic switching power losses can be expected with a polarization error in the QCA circuit. In this work, a temperature value of 2 K was taken as the QCAPro parameter. A comparative analysis of the dissipated power at different levels of tunneling energy (0.5 $E_k$, 1 $E_k$, and 1.5 $E_k$) for the proposed flip-flop is shown in Table 5. The maps of the dissipated power for the presented flip-flop with a tunneling energy of 0.5 $E_k$ are illustrated in Figure 12.

Table 5. Power dissipation results.

| Circuit Presented | Average of Leakage Energy Dissipation (meV) | Average of Switching Energy Dissipation (meV) | Total Energy Consumption (meV) |
|---|---|---|---|

| in | 0.5Ek | 1Ek | 1.5Ek | 0.5Ek | 1Ek | 1.5Ek | 0.5Ek | 1Ek | 1.5Ek |
|---|---|---|---|---|---|---|---|---|---|
| [21] | 45.67 | 131.9 | 231.18 | 90.75 | 79.16 | 67.86 | 136.42 | 211.06 | 299.04 |
| [27] | 19.23 | 53.9 | 93.58 | 36.23 | 30.76 | 25.83 | 55.46 | 84.66 | 119.41 |
| [22] | 22.48 | 67.46 | 120.46 | 66.24 | 58.25 | 50.13 | 88.72 | 125.71 | 170.59 |
| [10] | 15.76 | 44.91 | 78.23 | 15.6 | 13.49 | 11.49 | 31.36 | 58.4 | 89.72 |
| proposed | 6.27 | 18.49 | 32.67 | 22.94 | 19.72 | 16.73 | 29.22 | 38.22 | 49.40 |

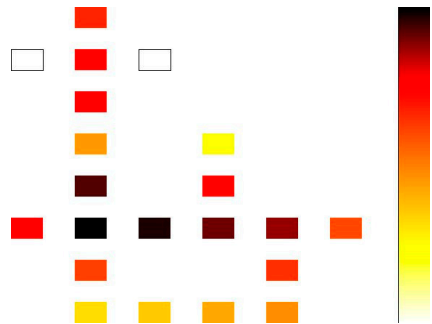

**Figure 12.** Power dissipation maps for the proposed LST-FF with a tunneling energy level of $0.5E_k$ at a temperature of 2 Kelvin.

## 7. Conclusions

This paper introduced an optimal form of the T flip-flop. The presented flip-flop was level-sensitive and was implemented with a noticeable area of 0.0186 μm$^2$ and at minimum complexity with only 21 cells. The unique proposed design of the flip-flop was utilized to implement synchronous counters with different sizes. However, only the 3-bit counter was reviewed in detail. This counter had the lowest number of cells and the smallest area. The verification of the proposed circuits was performed using the QCADesigner software. A suitable converter with an optimal structure was designed to provide the counter with the ability to detect the edges of the clock signal. Another important feature in the proposed design compared to others was that all the outputs were in the terminals of the circuit.

**Author Contributions:**

Conceptualization, Ali H.Majeed; Methodology, Esam Alkaldy; Supervision, Mohd Shamian bin Zainal and Danial Bin MD Nor; Validation, Ali Majeed; Writing—original draft, Ali Majeed; Writing—review & editing, Esam Alkaldy.

**Funding:** This research received no external funding.

**Conflicts of Interest**: The authors declare no conflict of interest.